\documentclass[sigconf]{acmart}
\AtBeginDocument{%
  }

\copyrightyear{2026}
\acmYear{2026}
\setcopyright{cc}
\setcctype{by}
\acmConference[WWW '26]{Proceedings of the ACM Web Conference 2026}{April 13--17, 2026}{Dubai, United Arab Emirates}
\acmBooktitle{Proceedings of the ACM Web Conference 2026 (WWW '26), April 13--17, 2026, Dubai, United Arab Emirates}
\acmPrice{}
\acmDOI{10.1145/3774904.3792628}
\acmISBN{979-8-4007-2307-0/2026/04}

\usepackage{subcaption}
\usepackage{pgfplots}
\usepackage{xcolor}
\usepackage{tikz}
\usepackage{amsmath}
\usepackage{enumitem}
\usepackage{bm}
\usepackage{threeparttable}
\usepackage{graphicx}
\usepackage{multirow}
\usepackage{pgf-pie}
\usepackage{tabularx}
\usepackage{appendix}
\usepackage[nameinlink]{cleveref}

\pgfplotsset{compat=newest}
\usetikzlibrary{plotmarks}

\begin{document}

\title{Modeling Stage-wise Evolution of User Interests for News Recommendation}

\author{Zhiyong Cheng}
\affiliation{%
  \institution{The School of Computer Science and Information Engineering, Hefei University of Technology}
  \city{Anhui}
  \country{China}
  }
\email{jason.zy.cheng@gmail.com}

\author{Yike Jin}
\affiliation{%
  \institution{The School of Computer Science and Information Engineering, Hefei University of Technology}
  \city{Anhui}
  \country{China}
  }
\email{jyk06801@gmail.com}

\author{Zhijie Zhang}
\affiliation{%
  \institution{The School of Computer Science and Information Engineering, Hefei University of Technology}
  \city{Anhui}
  \country{China}
  }
\email{zhijiezhang021@gmail.com}

\author{Huilin Chen}
\affiliation{%
  \institution{The School of Computer Science and Information Engineering, Hefei University of Technology}
  \city{Anhui}
  \country{China}
  }
\email{ClownClumsy@outlook.com}

\author{Zhangling Duan}
\authornote{Zhangling Duan is the corresponding author.}
\affiliation{%
  \institution{Institute of Artificial intelligence, Hefei Comprehensive National Science Center}
  \city{Anhui}
  \country{China}
  }
\email{duanzl1024@iai.ustc.edu.cn}

\author{Meng Wang}
\affiliation{%
    \institution{The School of Computer Science and Information Engineering, Hefei University of Technology}
  \city{Anhui}
  \country{China}
  }
\email{eric.mengwang@gmail.com}

\begin{abstract}
Personalized news recommendation is highly time-sensitive, as user interests are often driven by emerging events, trending topics, and shifting real-world contexts. These dynamics make it essential to model not only users’ long-term preferences, which reflect stable reading habits and high-order collaborative patterns, but also their short-term, context-dependent interests that change rapidly over time. However, most existing approaches rely on a single static interaction graph, which struggles to capture both long-term preference patterns and short-term interest changes as user behavior evolves.
To address this challenge, we propose a unified framework that learns user preferences from both global and local temporal perspectives. A global preference modeling component captures long-term collaborative signals from the overall interaction graph, while a local preference modeling component partitions historical interactions into stage-wise temporal subgraphs to represent short-term dynamics. Within this module, an LSTM branch models the progressive evolution of recent interests, and a self-attention branch captures long-range temporal dependencies. Extensive experiments on two large-scale real-world datasets show that our approach consistently outperforms strong baselines and delivers fresher and more relevant recommendations across diverse user behaviors and temporal settings.
\end{abstract}



\maketitle

\section{Introduction}
Recommender systems~\cite{qi2022caum,garrido2024sjors,cai2024mitigating,chen2025i3,cai2026} have become indispensable for alleviating information overload and are widely applied in domains such as e-commerce, video, music, and news~\cite{ozgobek2014survey,li2019survey,raza2022news,meng2023survey,iana2024survey,moller2025explaining}. In the news domain, the challenge is particularly severe: Online platforms generate a massive stream of news every day, making it challenging for users to filter through the flood of content and identify what truly matters to them. Personalized news recommendation has thus become essential for delivering timely and relevant information.  

Compared with other domains, news recommendation faces unique challenges due to the \textit{time-sensitive} nature of news content. The relevance of articles quickly loses recommendation value as new events occur, or trending topics emerge. Users generally seek the most up-to-date information and lose interest in outdated reports. Moreover, user preferences are not static but exhibit temporal fluctuations driven by real-world events and seasonal contexts. For example, sports fans may focus on major tournaments like the World Cup, while health-related news consumption may spike during a pandemic. Consequently, recommendation systems must account for both \textit{long-term interests} reflecting stable reading habits, and \textit{short-term dynamics } capturing rapid preference shifts triggered by emerging events. How to effectively preserve stable interests while adapting to rapidly changing ones remains a fundamental issue to be addressed in personalized news recommendation.

Early news recommendation systems followed two  paradigms: collaborative filtering (CF) and content-based (CB) methods. CF approaches~\cite{resnick1994grouplens,das2007google,ahn2008new} infer preferences from behavioral similarity, while CB approaches~\cite{lang1995newsweeder,gabrilovich2004text,li2010contextual,liu2010personalized} leverage textual content, categories, and metadata to model relevance. Hybrid methods~\cite{claypool1999combining,wang2011collaborative} combine both signals to alleviate sparsity and cold-start issues. 
With the rise of deep learning, neural recommenders significantly improved representation learning by capturing richer semantics. 
Early neural models such as DSSM~\cite{huang2013learning} demonstrated the effectiveness of deep semantic representation learning for content understanding.
Subsequently, representative news recommenders including NPA~\cite{wu2019npa}, NRMS~\cite{wu2019neural}, and NAML~\cite{wu2019naml} enhanced user and item embeddings via attention mechanisms, while models such as FIM~\cite{wang2020fine} further captured fine-grained semantic interactions between user histories and candidate news.
Knowledge-aware approaches like DKN~\cite{wang2018dkn} integrated external knowledge graphs to enrich semantic representations.
Later works further emphasized structural and collaborative signals by employing GCN-based encoders over user–news graphs, leveraging high-order connectivity to improve representation consistency~\cite{mao2021neural,ge2020graph,hu2020graph}.
However, these models typically assume user preferences to be static and fail to capture how they evolve over time.

To better capture the temporal nature of news consumption, subsequent work introduced time-aware designs.
LSTUR~\cite{an2019neural} fuses a long-term user embedding with short-term sequential behaviors to capture both stability and recency, while HieRec~\cite{qi2021hierec} organizes interactions into hierarchical sub-sequences to model multi-scale temporal evolution.
More recent studies further pursue fine-grained preference modeling, including intent disentanglement~\cite{wu2021user,wu2022two,ko2025crown} and temporal contrastive strategies such as TCCM~\cite{chen2023tccm}, which differentiate user interests across time segments.

Despite these advancements, the challenge of balancing short-term sensitivity with long-term stability has received limited attention and remains largely underexplored in existing work. Models that rely on static global graphs cannot capture evolving interests and often produce outdated recommendations. On the other hand, sequential models that emphasize recent behaviors often fail to capture stable collaborative signals, leading to representations that are biased toward short-term interests. As a result, user profiles can become either dominated by outdated interactions or overly responsive to temporary trends, both of which reduce recommendation accuracy. This gap motivates the need for a unified approach that simultaneously leverages global collaborative structures and captures dynamic, time-dependent behaviors.

In this paper, we propose a unified framework that addresses this challenge by jointly modeling long-term preferences and short-term dynamics. Our key insight is that user–news interactions naturally exhibit a stage-wise temporal structure: clicks within the same period often reflect coherent local interests, while transitions across periods capture how preferences evolve over time.
To exploit this property, our framework consists of two components:
(1) A \textbf{global preference modeling } module that learns stable, high-order collaborative signals from the entire interaction graph.
(2) A \textbf{local preference modeling} module that partitions user histories into temporal subgraphs to capture stage-specific behaviors. Within this module, a LSTM captures the progressive evolution of short-term interests, while a self-attention mechanism aggregates informative signals across different stages to model long-range dependencies.
By combining global preference stability with local temporal dynamics, our approach achieves a better balance between the two, leading to more accurate and robust modeling of evolving user interests in news recommendation. 

To validate the effectiveness of our approach, we conduct extensive experiments on two real-world datasets. The results show that our model consistently outperforms state-of-the-art baselines and achieves more accurate modeling of evolving user preferences. Further ablation studies verify the contribution of each component. We have released the code and relevant parameter settings to facilitate repeatability as well as further research\footnote{https://github.com/yikeJin/SEIN.}.

In summary, our work makes the following key contributions:  
\begin{itemize}[leftmargin=*]
    \item We tackle the underexplored challenge of modeling stage-wise evolution of user interests in news recommendation, where preferences are shaped by both long-term reading habits and rapidly changing contextual events.  
    \item We propose a unified framework that jointly learns global collaborative patterns from the full interaction graph and local stage-specific dynamics from temporally segmented subgraphs. The latter enables fine-grained modeling of short-term interest evolution via LSTM-based transitions and self-attention–based cross-stage aggregation.  
    \item We demonstrate the effectiveness and robustness of our approach through extensive experiments on large-scale real-world datasets, with comprehensive analyses validating its advantages under diverse temporal granularities and user behaviors.  
\end{itemize}

\section{Related Work}
\label{sec:relatedwork}


\textbf{Traditonal methods.} Early approaches to news recommendation followed the paradigms of collaborative filtering (CF) and content-based (CB) recommendation. CF-based methods~\cite{resnick1994grouplens,das2007google,ahn2008new,liu2021interest,liu2025understanding} infer user preferences by leveraging behavioral similarity, assuming that users with similar interaction patterns are likely to consume similar news. Matrix factorization and its variants are widely used in this setting, but they often suffer from data sparsity and cold-start  issues. CB methods~\cite{lang1995newsweeder,gabrilovich2004text,ijntema2010ontology,li2010contextual,liu2010personalized,huang2013learning,wang2018dkn,liu2024cluster}, in contrast, directly exploit textual content, categories, and metadata to build richer news representations and improve recommendation for unseen items. Hybrid approaches~\cite{claypool1999combining,wang2011collaborative,li2011scene,de2012chatter,tarus2018hybrid} combine the strengths of both paradigms by integrating collaborative signals with semantic understanding.

\textbf{Neural news recommenders.} With the rise of deep learning, neural architectures have significantly advanced personalized news recommendation by learning richer semantic representations from textual and auxiliary information.
Representative works include DAN~\cite{zhu2019dan} and DKN~\cite{wang2018dkn}, which leverage attention or multi-channel convolutional networks to model news semantics and knowledge-level signals; NPA~\cite{wu2019npa}, which introduces personalized attention to highlight informative content; NRMS~\cite{wu2019neural}, which employs multi-head self-attention to capture contextual dependencies; and NAML~\cite{wu2019naml}, which fuses multiple views of news content (titles, abstracts, and categories) via hierarchical attention. HieRec~\cite{qi2021hierec}, which models user interests in a hierarchical structure to capture multi-level preference representations.
These methods greatly improve representation learning but largely treat interactions as independent events, overlooking structural and temporal relationships.

\textbf{GCN-based methods.} 
Graph neural networks (GCNs) have been widely adopted in news recommendation to exploit high-order connectivity and structural correlations in user–news interaction data. Early approaches applied GCNs to the bipartite user–news graph to enhance representation learning beyond direct interactions. 
For example, session-aware models~\cite{sheu2020context} capture short-term behavioral signals at the session level. GNewsRec~\cite{hu2020graph} further extends the bipartite structure into a heterogeneous user–news–topic graph to encode richer semantic relationships.
Subsequent work further expands the graph structure to encode more complex relations: DIGAT~\cite{mao2022digat} models dual-graph interactions over user–user and news–news links, and NRMG~\cite{chen2023nrmg} proposes a multiview GCN to integrate user–news, news–entity, and news–topic relations into a unified framework. More recently, HGNN-PAD~\cite{zhang2024heterogeneous} introduces a heterogeneous GNN with personalized adaptive diversity to balance accuracy and novelty in recommendations.
These methods demonstrate the effectiveness of structural modeling, yet they typically rely on a single static global graph, which limits their ability to capture how user preferences evolve over time.

\textbf{Temporal modeling.} Since user interests shift continuously with emerging events and changing contexts, temporal modeling has become a major research focus. Early attempts such as LSTUR~\cite{an2019neural} explicitly separate long-term and short-term preferences. FeedRec~\cite{wu2022feedrec} further leverages diverse feedback to dynamically construct user representations. Other approaches~\cite{cai2024popularity,cai2025graph,ferrara2024divan} incorporate temporal and causal signals to address bias or cold-start issues. 
For instance, TCCM~\cite{chen2023tccm} introduces a time- and content-aware causal model, HyperNews~\cite{liu2020hypernews} jointly models recommendation and active-time prediction, and DIVAN~\cite{ferrara2024divan} incorporates popularity- and virality-aware temporal signals for news recommendation.
Beyond news recommendation, dynamic graph-based methods such as DGSR~\cite{zhang2022dynamic} and TGCL4SR~\cite{zhang2024temporal} learn evolving collaborative signals by converting interaction sequences into dynamic graphs or performing temporal contrastive learning.

Despite these advances, most existing methods either rely solely on a static global user–news graph or focus exclusively on short-term sequential modeling. The former captures long-term collaborative signals but overlooks stage-specific behavioral changes, while the latter models temporal shifts but fails to incorporate stable, high-order structures. This gap highlights the need for a unified framework that integrates both perspectives. 
To address these limitations, we propose a recommendation framework that jointly models long-term user preferences and short-term interest dynamics. By combining global collaborative patterns with temporally stage-wise interaction signals, our approach captures how user interests evolve over time, providing a more adaptive and accurate solution for personalized news recommendation.
\section{Methodology}
\subsection{Preliminaries}

\label{sec:prelim}
In this paper, we focus on the news recommendation task, which aims to predict the likelihood that a target user will click on a candidate news article based on her historical interaction behavior. Formally, let $\mathcal{U} = \{u_1, u_2, \ldots, u_{|\mathcal{U}|}\}$ denote the set of users and $\mathcal{I} = \{i_1, i_2, \ldots, i_{|\mathcal{I}|}\}$ denote the set of news articles.
For a user $u \in \mathcal{U}$, her chronological click history can be represented as:
\begin{equation}
H_u = \{(i^u_1, \psi^u_1), (i^u_2, \psi^u_2), \ldots, (i^u_M, \psi^u_M)\},
\end{equation}
where $i^u_m \in \mathcal{I}_u$ is the $m$-th clicked news article with user $u$, $\psi^u_m \in \Psi_u$ denote the timestamp of the interaction $(u, i^u_m)$, and $M$ is the total number of clicks. Given a set of candidates $\mathcal{C}_u \subseteq \mathcal{I}$, the task is to estimate the probability that user $u$ will click on each news article $i \in \mathcal{C}_u$ that she has not previously interacted with.

To capture users’ time-evolving interests, we first convert the historical clicked timeline $\Psi=\{\psi_1,\psi_2,\ldots, \psi_{|M|}\}$ into a time-interval sequence $T=\{t_1, t_2, \ldots, t_{|T|}\}$, where the temporal intervals $\Delta t= |\psi_j - \psi_i|$. Note that this period can be defined at different granularities, such as weekly or monthly windows. Therefore, we convert the continuous time sequence into discrete time intervals, where each distinct period reflects coherent local interests.

In our task, we initialize the embedding vectors of a user $u \in \mathcal{U}$ and a news item $i \in \mathcal{I}$ as $\bm{e_u^{0}} \in \mathbb{R}^d$ and $\bm{e_i^{0}} \in \mathbb{R}^d$, where $d$ denotes the embedding dimension.
For users, we randomly initialize their embeddings from a trainable matrix $\bm{P} \in \mathbb{R}^{d \times  |\mathcal{U}| }$. Each user is represented by a unique one-hot ID vector $\bm{ID}_u^{U}$, and the initial embedding is computed as:
\begin{equation}
\bm{e_{u}^{0}} = \bm{P} \cdot \bm{ID}_u^{U}.
\end{equation}

For news items, we incorporate textual semantics into the initialization. Specifically, we use a pre-trained language model to encode each news title into a semantic vector and then project it into the $d$-dimensional latent space through a single-layer neural network:
\begin{equation}
\bm{e_{i}^{0}} = \mathrm{MLP}(\mathrm{LM}(\text{Title}_i)).
\end{equation}
This semantic initialization provides richer content-aware representations for news articles, while user embeddings are learned purely from interaction signals during training.

\subsection{Model}

\subsubsection{Overview.}
This work aims to model user interests with both \textit{long-term preferences} from historical behaviors and \textit{short-term interests} driven by emerging contexts. We develop a framework with two components:
\begin{itemize}[leftmargin=*]
\item \textbf{Global preference modeling {(GPM)}:} It captures long-term preferences, which reflect the relatively stable interests of the users accumulated over a long history of interactions (e.g., a user who consistently follows finance or sports news). Such preferences encode stable, high-order collaborative signals across the entire user–news interaction graph. By aggregating global structural information, the encoder builds comprehensive user representations that provide a strong foundation for subsequent recommendation and downstream preference modeling.

\begin{figure*}[t]
    \vspace{-0.2cm}
    \centering
    \includegraphics[width=1\textwidth]
    {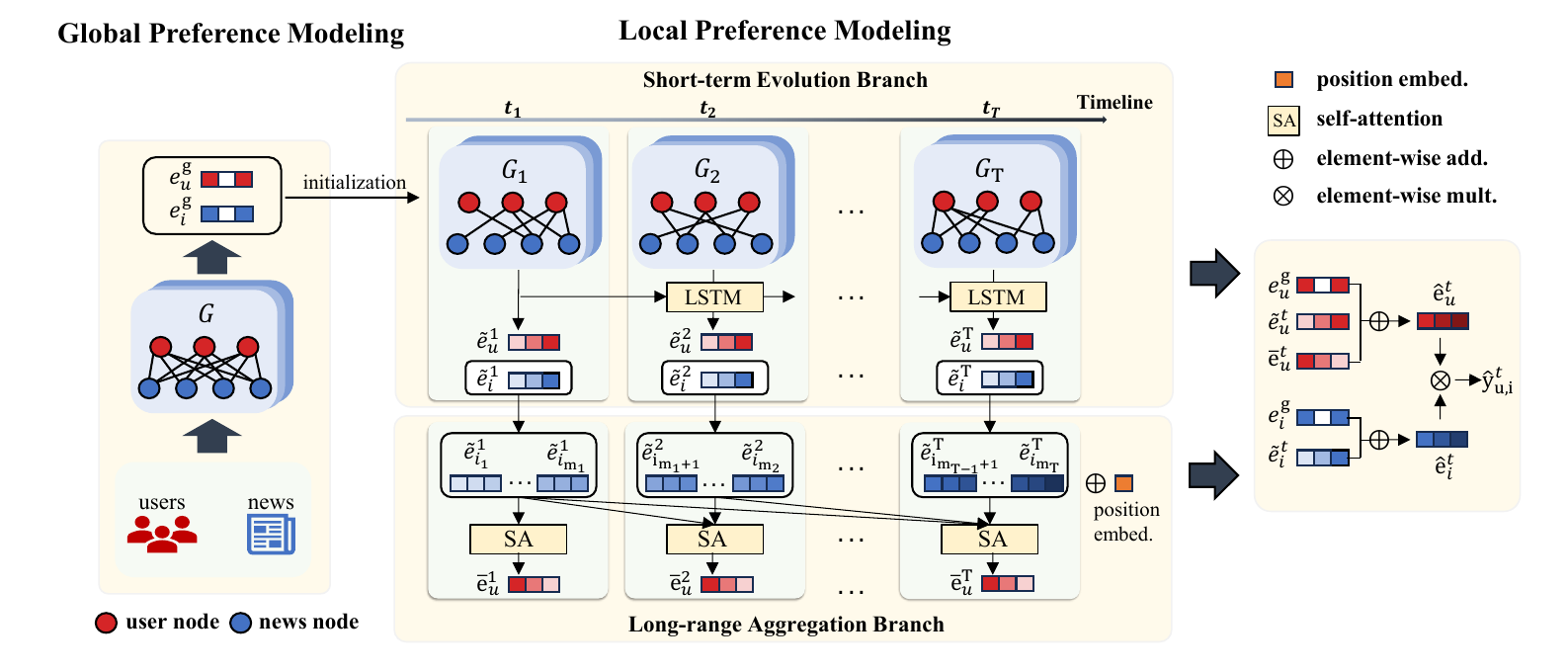}
    \vspace{-0.4cm}
    \caption{Overall framework of our model. It integrates a Global Preference Modeling module for stable collaborative preferences 
    and a Local Preference Modeling module that captures temporal dynamics through Short-term Evolution Branch and Long-range Aggregation Branch.}
    \label{fig:framework}
    \vspace{-0.4cm}
\end{figure*}

\item \textbf{Local preference modeling (LPM):} While global modeling captures stable behavioral patterns, user interests also change over time in response to new contexts. To address this, LPM partitions historical interactions into \textit{stage-wise temporal subgraphs} and models how preferences evolve across them, enabling the recommender to stay sensitive to short-term dynamics while preserving historical context.
\end{itemize}
By combining these two modules, our framework jointly captures global preference stability and temporal interest dynamics, leading to a more comprehensive representation of evolving user preferences for personalized news recommendation.

\begin{itemize}[leftmargin=*]
\item \textbf{Global Graph.}
We build a global user–news interaction graph $\mathcal{G} = (\mathcal{U} \cup \mathcal{I}, \mathcal{E})$, where $(u, i) \in \mathcal{E}$ if user $u$ clicked news $i$ at any point in time. This global graph encodes collaborative signals across the entire interaction history and serves as the basis for learning global representations.

\item \textbf{Temporal Subgraphs.}
We partition the chronological click history $\mathcal{I}_u$ into $T$ discretive time-interval periods, denoted as $\{\mathcal{I}_u^1, \mathcal{I}_u^2,\ldots,\mathcal{I}_u^T\}$. Each interaction $\mathcal{I}_u^t \subseteq \mathcal{I}_u$ represents a subset of interaction corresponding to the $t$-stage. Under this strategy, each time-interval interaction forms a temporal subgraph $\mathcal{G}_t = (\mathcal{U} \cup \mathcal{I}, \mathcal{E}_t)$, where $\mathcal{E}_t = \{(u,i) \mid u \text{ clicked } i \text{ during } t \text{ period } \}$. This segmentation prevents heterogeneous signals from being aggregated indiscriminately and preserves stage-specific characteristics.
\end{itemize}

\subsubsection{Global Preference Modeling}
\label{sec:global}
We capture users’ long-term preferences by learning from the overall user–news interaction graph, which encodes stable thematic interests and high-order collaborative signals (e.g., a user consistently reading finance or sports news). To this end, we adopt LightGCN~\cite{he2020lightgcn}, a widely used and efficient GCN-based recommendation model, to generate user and item representations from the global graph $\mathcal{G}$.

Formally, given the initial embeddings $\bm{e_u^{(0)}}$ and $\bm{e}_i^{(0)}$ for a user $u$ and an item $i$, the $l$-th layer propagation is defined as:
\begin{equation}
\bm{e}_u^{(l+1)} = \sum_{i \in \mathcal{N}_u}
\frac{1}{\sqrt{|\mathcal{N}_u|}\sqrt{|\mathcal{N}_i|}} \bm{e}_i^{(l)},
~~
\bm{e}_i^{(l+1)} = \sum_{u \in \mathcal{N}_i}
\frac{1}{\sqrt{|\mathcal{N}_u|}\sqrt{|\mathcal{N}_i|}} \bm{e}_u^{(l)},
\end{equation}
where $\mathcal{N}_u$ and $\mathcal{N}_i$ denote the neighbor sets of user $u$ and item $i$, respectively.
After $K$ layers of propagation, we obtain a sequence of intermediate embeddings for each user and item: $\{\bm{e}_u^{(0)}, \bm{e}_u^{(1)}, \ldots, \bm{e}_u^{(l)}\}$ and $\{\bm{e}_i^{(0)}, \bm{e}_i^{(1)}, \ldots, \bm{e}_i^{(l)}\}$. The final global embeddings are computed by simply aggregating these layer-wise representations:
\begin{equation}
\bm{e}_u^g = \sum_{l=0}^{L} \bm{e}_u^{(l)}, \quad
\bm{e}_i^g = \sum_{l=0}^{L} \bm{e}_i^{(l)}.
\end{equation}

The global embeddings $\bm{e_u^g}$ and $\bm{e_i^g}$ are used to initialize embedding learning in temporal subgraphs. Since interactions within each temporal segment are typically much sparser than those in the global graph, high-quality initialization plays a crucial role in learning good representations. Leveraging the globally learned embeddings as starting points provides richer prior knowledge and facilitates more effective embedding learning in the temporal modules, which has also been validated in our experiments.





\subsubsection{Local Preference Modeling}
\label{sec:etrl}
While global modeling captures stable collaborative signals, it fails to model temporal interest evolution. To address this, the LPM component divides chronological user–news interactions into temporal subgraphs, each reflecting preferences within a specific period, and models their evolution across stages. LPM contains two complementary branches: (1) a \textit{\textbf{short-term evolution branch}} using LSTM to capture progressive and recency-sensitive preference shifts; (2) a \textit{\textbf{long-range aggregation branch}} employing self-attention to model cross-stage dependencies and highlight historically influential signals. Together, these branches capture both short-term dynamics and long-range dependencies, producing richer representations of user interest evolution.

\textbf{Short-term Evolution Branch.}
User preferences often change rapidly in response to new events or trending topics, which makes capturing short-term dynamics critical for recommendation relevance. To achieve this, we learn user and item representations on each temporal subgraph $\mathrm{GCN}^{(t)}$ separately, conditioned on the global embeddings $(\bm{e_u^g}, \bm{e_i^g)}$:
\begin{equation}
\bm{e}_u^t, \bm{e}_i^t = \mathrm{GCN}^{(t)}\big(\mathcal{G}_t; \bm{e}_u^g, \bm{e}_i^g\big), 
~~ t=\{1,\ldots,T\}.
\end{equation}
This stage-specific representation prevents signals from distant periods from being indiscriminately mixed and preserves the semantics unique to each time window.  

To further capture the evolution of preferences across adjacent stages, we model sequential dependencies with an LSTM:
\begin{equation}
\bm{\tilde e}_u^t = \mathrm{LSTM}(\bm{e}_u^t, \bm{e}_u^{(t-1)}), \quad
\bm{\tilde e}_i^t = \mathrm{LSTM}(\bm{e}_i^t, \bm{e}_i^{(t-1)}),
\end{equation}

This recurrent design links temporal representations over time, enabling the model to track progressive preference shifts and remain sensitive to users’ most recent behaviors.

\textbf{Long-range Aggregation Branch.}
While the LSTM models local transitions, it may overlook the long-range interest evolution influenced by earlier interactions. Therefore, we introduce a self-attention–based aggregation mechanism that considers all temporal subgraphs within a unified representation space. 
By adaptively weighting historical signals alongside recent ones, this branch captures \textit{stage-wise evolution}. This long-range aggregation offers a global temporal perspective that complements the local dynamics modeled by the LSTM, leading to richer user representations and more accurate recommendations in dynamic news environments.

Consider a user $u$ with a chronological click history of $M$ items indexed by $\{i_1, i_2, \ldots, i_M\}$. The timeline is discretized into $T$ temporal intervals $\{t_1, t_2, \ldots, t_T\}$. For each stage $t$, we define the prefix of observed clicks up to the end of interval $t$ as $\{ i_1, i_2, \ldots, i_{m_t} \}$, where $0 < m_1 < m_2 < \cdots < m_T = M$. At stage $t$, we construct a stage-wise cumulative representation $\bm{S}_u^{t}$ by aggregating all historical items clicked before or within that stage:
\begin{equation}
\bm{S_u^{t}} = 
\big[\, 
\bm{\tilde e}_1^t + \bm{p}_1,\;
\bm{\tilde e}_2^t + \bm{p}_2,\;
\ldots,\;
\bm{\tilde e}_{m_t}^t+ \bm{p}_{m_t}
\,\big],
\end{equation}
where $\bm{\tilde e}_i^{t}$ denotes the embedding of clicked news item $i$ obtained the LSTM output at each stage.  $\bm{p}_j$ is a learnable positional encoding vector that serves as a chronological anchor.
This design enables the model to distinguish earlier from more recent clicks within each prefix and to adaptively assign stage-dependent importance as the interaction history accumulates over time.

To capture dependencies within the entire prefix, we stack $L_a$ layers of self-attention with residual connections:
\begin{equation}
\bm{S}_{u,l}^{(t)} \;=\; \sigma\!\big(\mathrm{Self\mbox{-}Att}(\bm{S}_{u,l-1}^{(t)})\big) \;+\; \bm{S}_{u,l-1}^{(t)}, 
~~ l=\{1,\ldots,L_a\},
\end{equation}
where $S_{u,0}^{(t)}=S_u^{(t)}$ and $\sigma(\cdot)$ is a softmax function.
The stage-$t$ cumulative long-term representation is then obtained by summing over positions after the last attention layer:
\begin{equation}
\bm{\overline{e}}_u^{t} \;=\; \sum\nolimits \bm{S}_{u,l}^{(t)}, 
~~ l=\{1,\ldots,L_a\}.
\end{equation}

Applying the above construction to all stages yields the time-indexed cumulative user representations: $\{\bm{\overline{e}}_u^{1},\,\bm{\overline{e}}_u^{2},\,\ldots,\,\bm{\overline{e}}_u^{T}\}$
where $\bm{\overline{e}}_u^{t}$ denotes the user representation up to $t$.

\textbf{Discussion.}
The two branches complement each other by modeling different aspects of user interest evolution. The \textbf{short-term evolution branch} focuses on local, stage-to-stage transitions, capturing recency-sensitive preference shifts as user interests respond to new content. In contrast, the \textbf{long-range aggregation branch} captures stage-wise contextual dependencies across the entire interaction history, highlighting how earlier preferences interact with newly emerging ones. Together, they provide a holistic view of user dynamics, enabling the model to better balance immediate interests with long-term behavioral signals.

\subsection{ Prediction and Model Optimization}

\label{sec:train}

\subsubsection{Prediction}
Our model captures user preferences from three complementary perspectives: the short-term evolution branch $\bm{\tilde{e}}_u^t$ models stage-to-stage dynamics, the long-range aggregation branch $\bm{\overline{e}}_u^t$ captures cross-stage dependencies, and the global modeling component $\bm{e}_u^g$ encodes overall behavioral patterns.
To generate unified representations, we fuse the stage-specific, aggregated, and global embeddings:
\begin{equation}
\bm{\hat{e}}_u^t = \bm{\tilde{e}}_u^t + \bm{\overline{e}}_u^t + \bm{e}_u^g,\qquad
\bm{\hat{e}}_i^t = \bm{\tilde{e}}_i^t + \bm{e}_i^g.
\end{equation}
This fusion jointly preserves recent and long-term preference signals, enabling more accurate interaction prediction:
\begin{equation}
\hat{y}_{u,i}^{t} = \sigma(\bm{\hat{e}}_u^{t} \cdot \bm{\hat{e}}_i^{t}).
\end{equation}
where $\sigma(\cdot)$ denotes the sigmoid activation function that normalizes the interaction score into $(0,1)$, with larger values indicating a higher probability of user–item interaction.

\subsubsection{Objective Function}

\textbf{Prediction loss.}  
We train the model by minimizing a binary cross-entropy loss over the observed interactions within each temporal segment $t$. Specifically, given the predicted click probability $\hat{y}_{u,i}^t$ and the ground-truth label $y_{u,i}^t$ (where $y_{u,i}^t = 1$ if user $u$ clicked news item $i$ at time $t$, and $0$ otherwise), the objective is defined as:
\begin{equation}
L_t = - \!\!\sum_{(u,i) \in \mathcal{D}_t} \!\! \Big[\, y_{u,i}^t \log \hat{y}_{u,i}^t + (1 - y_{u,i}^t) \log (1 - \hat{y}_{u,i}^t) \Big],
\end{equation}
where $\mathcal{D}_t$ denotes the set of positive and sampled negative interactions in segment $t$. This loss encourages the model to assign higher scores to true interactions while suppressing false positives, effectively optimizing the click likelihood in each temporal window.

\textbf{Consistency regularization.} While temporal modeling captures dynamic user interests, it is desirable that these representations remain consistent with the global preference patterns learned from the complete interaction graph. To enforce such alignment, we introduce a contrastive consistency objective.
For each temporal segment $t$, the temporal embedding $\bm{\tilde{\mathbf e}_{u}^{t}}$ is encouraged to stay close to its corresponding global embedding $\bm{ e_{u}^{g}}$ while being separated from embeddings of other users:
\begin{equation}
\label{eq:con_infonce}
L_{\mathrm{cl}}^{t} 
= - \log 
\frac{\exp\!\big( \bm{\tilde{e}_{u}^t} \cdot \bm{e_{u}^{\,g}} / \tau \big)}
{\sum_{j} \exp\!\big( \bm{\tilde{e}_{u}^t} \cdot \bm{e_{j}^{\,g}} / \tau \big)} ,
\qquad
L_{\mathrm{cl}} = \sum_{t=1}^{T} L_{\mathrm{cl}}^{t} .
\end{equation}
Here, $\tau > 0$ is a temperature hyperparameter. This objective serves as a regularizer, improving representation robustness and bridging the gap between global and temporal modeling.

\textbf{Smoothness regularization.} Since user interests evolve gradually rather than abruptly, we further regularize the model by enforcing smooth transitions between adjacent temporal embeddings. A first-order smoothness constraint is applied:
\begin{equation}
\label{eq:smooth}
L_{\mathrm{sl}}^{t} 
= \big\| \bm{\tilde{e}_{u}^{\,t}}- \bm{\tilde{e}_{u}^{\,(t-1)}} \big\|_2^2 ,
\qquad
L_{\mathrm{sl}} = \sum_{t=1}^{T} L_{\mathrm{sl}}^{t} .
\end{equation}
This regularization mitigates noisy fluctuations and improves the model’s ability to capture the evolution of continuous preferences.

The final optimization objective integrates the above components, balancing the main recommendation signal with regularization terms.

\begin{equation}
\label{eq:full_obj}
L 
= \sum_{t=1}^{T}\lambda_t L_t 
+ \lambda_{\mathrm{cl}} \, L_{\mathrm{cl}}
+ \lambda_{\mathrm{sl}} \, L_{\mathrm{sl}}
+ \beta \, \|\Theta\|_2^2 .
\end{equation}
 The hyperparameters $\lambda_t$, $\lambda_{\mathrm{cl}}$, and $\lambda_{\mathrm{sl}}$ control the strengths of the main recommendation loss, consistency regularization, and smoothness regularization terms, respectively. The last term $\beta \|\Theta\|_2^2$ is a standard $\ell_2$ regularization applied to all trainable parameters $\Theta$, which helps prevent overfitting and improves the generalization ability of the model.

\section{Experiments}
\label{sec:experiments}

\noindent To evaluate our approach, we design experiments to address the following research questions:
\begin{enumerate}[leftmargin=*, label=\textbf{RQ\arabic*:}]
\item \textbf{Overall Performance:} Does our proposed model outperform state-of-the-art baselines ?
\item \textbf{Component Contribution:} How do individual components of our model contribute to the overall performance ?
\item \textbf{Hyperparameter Sensitivity:} How do different hyperparameters affect our model’s performance ?
\item \textbf{Model Capability:} Can our model effectively capture such periodic dynamics ?
\end{enumerate}

\subsection{Experimental Setup}
\subsubsection{Datasets and Dataset Split.}
We evaluate our method on two widely used news recommendation datasets: \textbf{Adressa}~\cite{gulla2017adressa} and \textbf{MIND}~\cite{wu2020mind}. Table~\ref{tab:dataset-statistics} summarizes the dataset statistics used in our experiments, including the number of users, items, interactions, and the time span covered. 

As user–news interactions naturally exhibit a stage-wise temporal structure, we partition each user’s click history into several discrete time intervals, where the interactions within each interval constitute a temporal subgraph. For the Adressa dataset, which provides complete timestamps in the official release, we adopt a chronological split: the last week is used for testing, the second-to-last week for validation, and the remaining data for training. For the MIND dataset, the official release contains week-6 data without labels, which is excluded from evaluation~\cite{iana2024survey,moller2025explaining}. According to the dataset organization, weeks 1–4 are chronologically ordered but do not provide exact timestamps; therefore, we divide them evenly into four pseudo-periods for training and validation, while week 5 is used as the test set.

\subsubsection{Implementation Details}
User embeddings are randomly initialized to 64 dimensions from a Gaussian distribution. We adopt Adam optimizer with a learning rate of $5 \times 10^{-6}$, batch size of 1024, embedding dimension of 64, and dropout rate of 0.2. For each positive click, we randomly sample four negative instances and apply early stopping based on validation AUC.  The temperature parameter is set to $\tau=0.1$. The coefficients in the loss function are empirically set as $\lambda_t=1e-1$, consistency regularization weight $\lambda_{cl}=1e-2$, and smoothness regularization weight $\lambda_{sl}=1e-2$. 

For temporal segmentation, we choose different window sizes based on the dataset characteristics. For the \textbf{Adressa-Large} dataset, which spans three months of interaction logs, the window size is set to \textbf{two weeks}. For the \textbf{MIND-Large} dataset, which contains only six weeks of logs, we set the window size to \textbf{one week} to ensure sufficient data for validation and testing. Unless otherwise specified, all experimental results reported in the following sections use these default segmentation settings.
 
\subsubsection{Baselines and Evaluation Metrics.}
To evaluate the effectiveness of our proposed approach, we compare it with a set of competitive baselines covering both classical and state-of-the-art news recommendation methods, as well as representative graph-based and temporal-aware models. (NRMS~\cite{wu2019neural}, NAML~\cite{wu2019naml}, NPA~\cite{wu2019npa}, LightGCN ~\cite{he2020lightgcn}, CNE-SUE~\cite{mao2021neural}, TCCM~\cite{chen2023tccm}, CROWN ~\cite{ko2025crown}).

For evaluation, we follow standard practice in the news recommendation field and adopt the area under the ROC curve (AUC), mean reciprocal rank (MRR), and normalized discounted cumulative gain (nDCG) as ranking metrics. Specifically, we report AUC, MRR, nDCG@5, and nDCG@10 on the testing set when the validation AUC is maximized.  All models are implemented in PyTorch~\footnote{https://pytorch.org/} with the Adam optimizer~\cite{kingma2014adam}, and all results are averaged over five independent runs to ensure robustness. Further details on the experimental protocol, dataset descriptions, hyperparameter settings, and implementation details
are provided in Appendix~\ref{app:dataset}.

\begin{table}[t]
\vspace{-0.3cm}
\centering
\caption{Statistics of news article datasets.}
\vspace{-0.3cm}
\label{tab:dataset-statistics}
\begin{tabular}{lcccc}
\toprule
\textbf{Dataset} & \textbf{Users} & \textbf{News} & \textbf{Clicks} & \textbf{Time Span} \\
\midrule
Adressa-large & 3{,}614{,}911 & 81{,}018 & 35{,}244{,}078 & 3 months \\
MIND-large & 1{,}000{,}000 & 161{,}013 & 24{,}155{,}470 & 6 weeks \\
\bottomrule
\end{tabular}
\vspace{-0.3cm}
\end{table}

\begin{table*}[t]
    \caption{Performance comparison of our model and baseline methods on the Adressa and MIND datasets. 
The best results are highlighted in \textbf{bold}, and the second-best results are \underline{underlined}.}
    \centering
    \resizebox{0.9\textwidth}{!}{
    \begin{tabular}{c|l|c|c|c|c|c|c|c|c}
    \hline
    Dataset & Metric & NPA & NRMS & NAML & LightGCN & CNE-SUE & TCCM & CROWN & Ours \\
    \hline
    \multirow{4}{*}{Adressa}
      & AUC     & 0.5312 & 0.5091 & 0.5046 & 0.4996 & 0.5207 & 0.5419 & \underline{0.6553} & \textbf{0.7993} \\
      & MRR     & 0.2521 & 0.2512 & 0.2602 & 0.4056 & 0.2471 & 0.3424 & \underline{0.3915} & \textbf{0.5223} \\
      & NDCG@5  & 0.2263 & 0.2206 & 0.2341 & 0.2812 & 0.2209 & 0.2380 & \underline{0.4156} & \textbf{0.4495} \\
      & NDCG@10 & 0.3103 & 0.3018 & 0.3002 & 0.3379 & 0.3046 & 0.3219 & \underline{0.5283} & \textbf{0.5327} \\
    \hline
    \multirow{4}{*}{MIND}
      & AUC     & 0.5008 & 0.5429 & 0.5479 & 0.5029 & 0.5266 & 0.5468 & \underline{0.5501} & \textbf{0.5804} \\
      & MRR     & 0.2947 & 0.3227 & 0.3173 & 0.2491 & 0.2366 & 0.3570 & \underline{0.3431} & \textbf{0.4368} \\
      & NDCG@5  & 0.3843 & 0.3566 & 0.3489 & 0.2289 & 0.2458 & 0.3178 & \underline{0.4156} & \textbf{0.4429} \\
      & NDCG@10 & 0.4176 & 0.4218 & 0.4134 & 0.2908 & 0.3077 & 0.4068 & \underline{0.4691} & \textbf{0.5171} \\
    \hline
    \end{tabular}
    } 
    \label{tab:adressalarge_mindlarge_results}
\end{table*}

\subsection{RQ1. Overall Performance}
Table~\ref{tab:adressalarge_mindlarge_results} reports the overall performance of our model compared with competitive baselines on the Adressa and MIND datasets. The best results are highlighted in bold, and the second-best results are underlined. Our approach consistently achieves substantial improvements over all baselines. These results demonstrate the effectiveness and robustness of our framework across datasets with different temporal spans and interaction characteristics.

Existing methods show clear limitations when handling dynamic, time-sensitive user preferences. Early attentive models such as NPA, NRMS, and NAML improve semantic representations but fail to model temporal evolution, leading to weaker ranking accuracy. 
Graph-based methods such as LightGCN and CNE-SUE capture high-order structural signals but rely on static graphs and fail to model evolving preferences.
Although TCCM considers temporal factors and alleviates popularity bias, its focus on global debiasing rather than fine-grained preference dynamics restricts its ranking performance. CROWN enhances disentangled and causal representation learning and achieves the second-best performance by exploiting content and category semantics, though its dependence on explicit content features restricts adaptability.

Our superior performance benefits from the complementary design of \textit{global} and \textit{local} preference modeling. The global preference encoder learns users’ long-term interests from their full interaction history and provides initialization for downstream temporal modeling, ensuring that local dynamics are grounded in a comprehensive behavioral context. In parallel, the local preference modeling captures evolving, context-dependent interests by constructing temporal subgraphs and modeling their sequential evolution. This combination allows the model to balance historical continuity with short-term adaptability, enabling it to track preference shifts while maintaining global consistency.


Overall, the integration of global preference modeling with fine-grained temporal dynamics provides a principled and effective solution for capturing evolving user interests.

\vspace{-0.2cm}
\subsection{RQ2. Ablation Study}
To quantify the contribution of each component in our framework, we conduct an ablation study on the Adressa-Large dataset by systematically removing key modules. The following model variants are evaluated:

\begin{itemize}[leftmargin=*]
    \item \textbf{w/o LPM}: Removes both the short-term evolution branch and the long-range aggregation branch (namely the \textit{local preference modeling} module), reducing the model to a basic version without temporal sequence modeling.
    \item \textbf{w/o STE}: Removes the \textit{short-term evolution} branch, leaving only the long-term aggregation pathway in LPM.
    \item \textbf{w/o LRA}: Removes the \textit{long-range aggregation} branch, retaining only the short-term evolution mechanism in LPM.
    \item \textbf{w/o GPM}: Removes the global preference modeling component, forcing the model to rely solely on local temporal dynamics.
\end{itemize}

The results are reported in Table~\ref{tab:ablation}. Several key observations can be made. First, the notable performance degradation of \textbf{w/o LPM} underscores the necessity of explicitly modeling temporal dynamics, as static global signals alone fail to capture evolving user behaviors.  

Second, comparing \textbf{w/o STE} and \textbf{w/o LRA} reveals a nuanced interplay between short-term evolution and long-range aggregation: removing the short-term branch causes a larger drop on metrics sensitive to recency (e.g., MRR and nDCG@5), highlighting its importance in adapting to rapidly shifting news interests. Conversely, the decline observed in \textbf{w/o LRA} demonstrates that merely tracking immediate changes is insufficient, incorporating non-local historical context significantly enhances the model’s understanding of user intent over longer horizons. 

Finally, the performance gap between the full model and \textbf{w/o GPM} illustrates the critical role of global preference modeling. Beyond capturing stable, high-order collaborative patterns, the global component also provides informative initialization for learning on sparse temporal subgraphs, which substantially improves convergence and representation quality. 

Together, these findings confirm that the three components: global preference modeling, short-term evolution, and long-range aggregation, jointly capture complementary aspects of user behavior. Their combination enables the model to track evolving preferences while grounding predictions in broader behavioral history, leading to consistently superior recommendation accuracy.

\begin{table}[t]
\vspace{-0.1cm}
\centering
\caption{Ablation study results on Adressa.}
\vspace{-0.2cm}
\label{tab:ablation}
\begin{tabular}{lcccc}
\toprule
Variant & AUC & MRR & N@5 & N@10 \\
\midrule
w/o LPM & 0.4880 & 0.1885 & 0.1049 & 0.1416 \\
w/o STE & 0.5109 & 0.2112 & 0.2570 & 0.2713 \\
w/o LRA & 0.4851 & 0.2573 & 0.3120 & 0.3322 \\
w/o GPM & 0.4996 & 0.4056 & 0.2812 & 0.3379 \\
\textbf{Ours} & \textbf{0.7993} & \textbf{0.5223} & \textbf{0.4495} & \textbf{0.5327} \\
\bottomrule
\end{tabular}
\vspace{-0.4cm}
\end{table}

\subsection{RQ3. Hyperparameter Sensitivity}
\begin{figure}[t]
    \centering
    \includegraphics[width=0.32\linewidth]{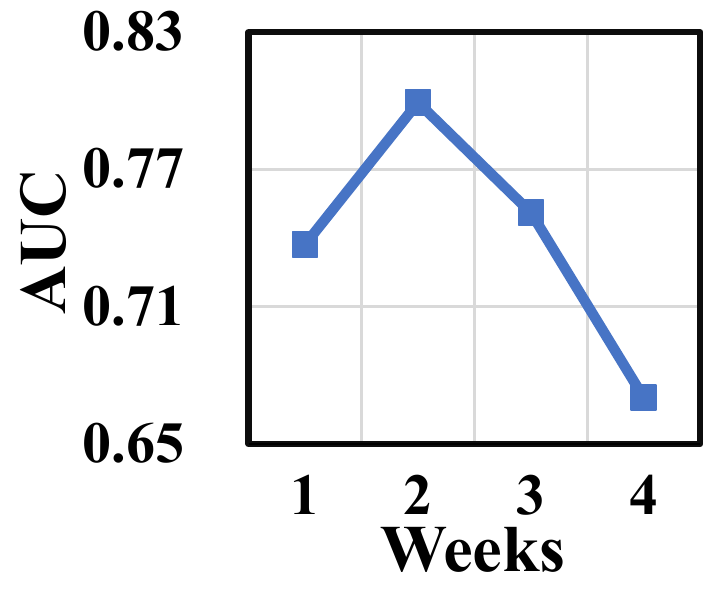}
    \includegraphics[width=0.32\linewidth]{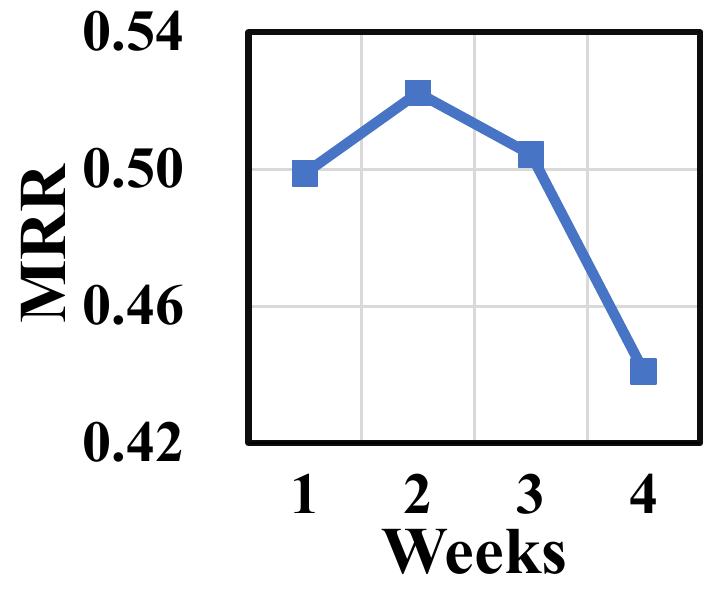}
    \includegraphics[width=0.32\linewidth]{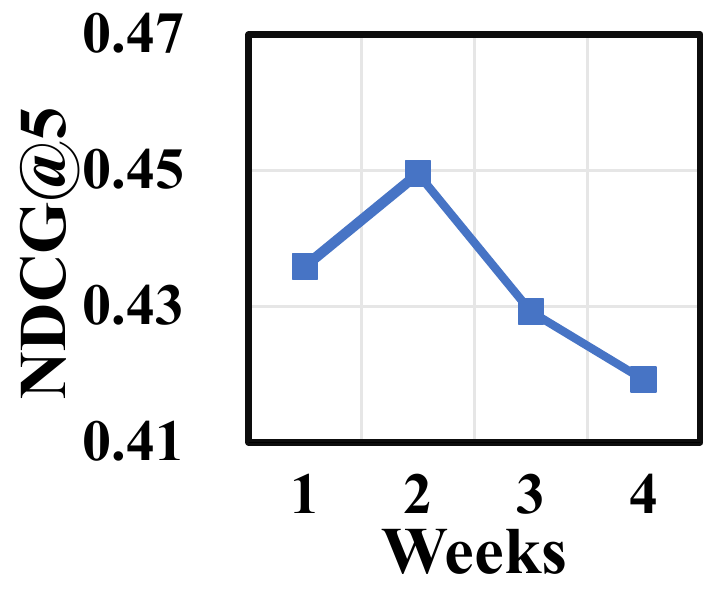}
    \vspace{-0.4cm}
    \caption{Performance under different window size.}
    \label{fig:win_gran}
    \vspace{-0.3cm}
\end{figure}

\textbf{Effects of Temporal Window Size.} 
The temporal segmentation window size is a crucial hyperparameter, as it determines the granularity of temporal subgraphs and thus affects how effectively the model captures preference dynamics. Very short windows may cause severe sparsity and unstable representations, while overly long windows can smooth out fine-grained interest shifts. To analyze its impact, we conduct a sensitivity study with window sizes of 1, 2, 3, and 4 weeks. Note that this analysis cannot be performed on MIND, since weeks 1–4 provide only chronological order without exact timestamps, making it impossible to construct temporal subgraphs under different window granularities. Therefore, we perform this experiment on the \textit{Adressa} dataset, which contains real-world timestamps, and report the results in Figure~\ref{fig:win_gran}. The results indicate that a 2-week window achieves the best overall performance in terms of AUC, MRR, and nDCG, suggesting that moderate segmentation strikes a good balance between modeling meaningful temporal dynamics and maintaining sufficient interaction density for representation learning.



\begin{figure}[t]
    \centering
    \vspace{-0.3cm}
    \subfloat[Prediction $\lambda_{t}$]{\includegraphics[width=0.32\linewidth]{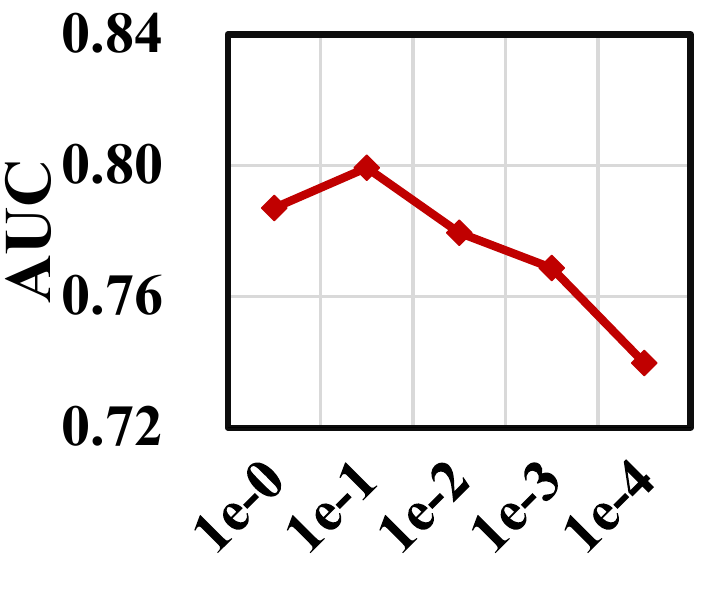}}
    \subfloat[Consistency $\lambda_{cl}$]{\includegraphics[width=0.32\linewidth]{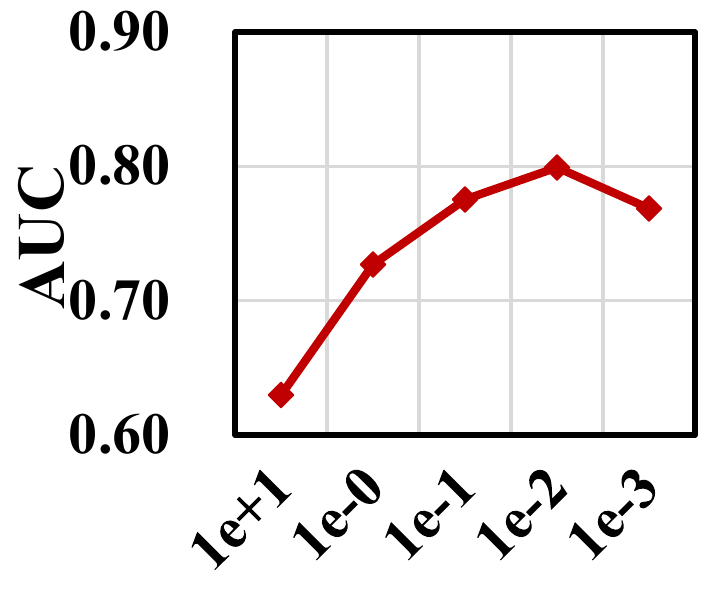}}
    \subfloat[Smoothness $\lambda_{sl}$]{\includegraphics[width=0.32\linewidth]{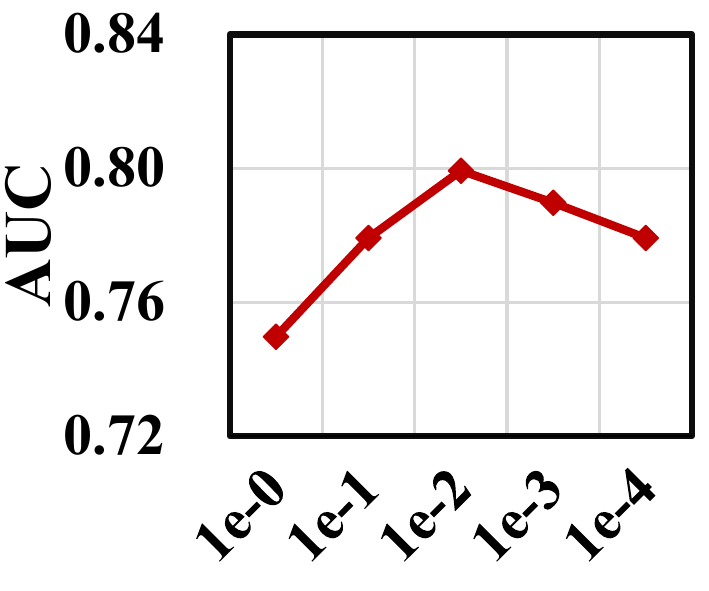}}
    \vspace{-0.4cm}
    \caption{Effect of loss weight coefficients on AUC.}
    \label{fig:coef_sensitivity}
    \vspace{-0.3cm}
\end{figure}

\textbf{Effects of Loss Weight Coefficients.}
We further investigate the effect of different coefficients in the objective function, including the weights for prediction, consistency, and smoothness terms. 
Here, we report the results on the \textbf{Adressa} dataset using \textbf{AUC} as the evaluation metric to provide a clear view of how each coefficient influences recommendation performance (Figure~\ref{fig:coef_sensitivity}), as the results for other metrics follow similar trends.

The \textit{prediction weight} achieves the highest AUC at $\lambda_{t} = 1e^{-1}$, indicating that assigning sufficient importance to the main objective is essential for stable convergence; in contrast, overly small $\lambda_{t}$ leads to underfitting on the primary recommendation task, resulting in weak temporal alignment and degraded preference modeling. For the \textit{consistency coefficient} $\lambda_{cl}$, the best performance is obtained at $1e^{-2}$, where moderate regularization encourages alignment between temporal and global embeddings, reduces semantic drift over time, and still preserves adaptability to short-term shifts. Similarly, the \textit{smoothness coefficient} $\lambda_{sl}$ reaches its optimum at $1e^{-2}$, effectively capturing gradual preference transitions without excessive smoothing—smaller values fail to maintain temporal continuity, whereas larger ones oversmooth representations and weaken the discrimination of evolving preferences.




\subsection{RQ4. Model Capability}
\label{sec:eq5}
To further evaluate whether our model can effectively capture the temporal evolution of user interests, we conduct an auxiliary experiment on the \textit{Adressa} dataset. 
Specifically, we randomly sample 20,000 users for each of four groups based on their interaction frequency: 
\emph{high-interaction} (avg.\ 377.8 clicks), \emph{medium-interaction} (avg.\ 39.6), 
\emph{low-interaction} (avg.\ 9.7), and a \emph{random} group drawn from the full population (avg.\ 86.2).

For evaluation, we categorize recommended news into two temporal groups:  
\textbf{new items} refer to articles published within the most recent one-month period of the dataset,  
while \textbf{historical items} refer to those published during the earliest one-month period.  
This distinction enables us to analyze how well the model adapts to emerging, time-sensitive content versus older content.

We compare the performance of the full model (with LPM) against a variant that uses only global preference modeling (\textit{w/o LPM}). 
Two metrics are reported:  
\textbf{New\%} and \textbf{Historical\%} indicate the proportion of new and historical articles in the Top-10 recommendations, respectively.  
\textbf{NRank} and \textbf{ORank} measure the average rank positions of new and historical items within the Top-10 list (lower values indicate higher ranking positions).

\begin{table}[t]
\centering
\caption{Comparison of models with and without LPM across user groups on Adressa. 
We report the proportion (\%) and average ranking positions of new and historical items in the Top-10 recommendations.}
\vspace{-0.2cm}
\label{tab:aux-etrl}
\begin{tabular}{l|c|c|c|c|c}
\toprule
Group & Model & New\% & Historical\% & NRank & ORank \\
\midrule
High  & w/o LPM & 0.035 & 0.036 & 1.59 & 1.58 \\
      & our model  & 0.014 & 0.018 & 0.60 & 0.82 \\
\midrule
Med   & w/o LPM & 0.164 & 0.161 & 4.31 & 4.32 \\
      & our model  & 0.187 & 0.145 & 4.53 & 3.85 \\
\midrule
Low   & w/o LPM & 0.159 & 0.156 & 4.10 & 4.12 \\
      & our model  & 0.173 & 0.148 & 4.09 & 3.74 \\
\midrule
Rand  & w/o LPM & 0.114 & 0.113 & 3.38 & 3.40 \\
      & our model  & 0.122 & 0.102 & 3.28 & 3.04 \\
\bottomrule
\end{tabular}
\vspace{-0.6cm}
\end{table}

\textbf{Observations.} The results in Table~\ref{tab:aux-etrl} lead to several important observations:

\begin{itemize}[leftmargin=*]
    \item \textbf{High-interaction users:} The full model significantly improves the ranking position of new articles (NRank~0.60 vs.\ 1.59), indicating that LPM leverages rich historical behavior to prioritize emerging, time-relevant content.
    \item \textbf{Medium- and low-interaction users:} LPM increases the share of new items (e.g., 18.7\% vs.\ 16.4\% for medium users) while improving the ranking of historical articles (3.85 vs.\ 4.32), suggesting a better balance between novelty and relevance.
    \item \textbf{Random user group:} Across a heterogeneous user population, LPM consistently improves both the proportion (12.2\% vs.\ 11.4\%) and ranking (3.28 vs.\ 3.38) of new items, demonstrating its robustness.
\end{itemize}

Overall, these results demonstrate that the LPM module substantially enhances the model’s ability to capture the temporal dynamics of user preferences. 
By adaptively prioritizing new and context-relevant news while still maintaining awareness of historically important content, our model produces recommendations that are both \textit{timely} and \textit{responsive} across diverse user groups.

\section{Conclusion}
\label{sec:conclusion}
We proposed a temporal interest subgraph modeling framework that jointly learns user preferences from both global and temporal perspectives. Specifically, the framework captures long-term collaborative patterns from the overall user–news interaction graph and leverages them as informative initialization for subsequent modeling. It then constructs temporal subgraphs to characterize stage-specific preferences, where a recurrent branch captures progressive short-term evolution, and a self-attention branch aggregates long-range dependencies across time. The framework addresses the limitations of static global graphs and short-term sequential models by integrating both long-term and short-term user interests.

\section{Acknowledgments}
This research is supported by the National Natural Science Foundation
of China and the New Cornerstone Science Foundation through the XPLORER PRIZE (Nos. 62272254, 72188101, 62020106007). Any opinions, findings, and conclusions or recommendations expressed in this material are those of the author(s) and do not reflect the views of the National Natural Science Foundation of China.

\bibliographystyle{ACM-Reference-Format}
\bibliography{sample-sigconf}

\appendix

\section{Detailed Experimental Settings}
\label{app:dataset}

\subsection{Datasets}

We provide detailed descriptions of the two publicly available datasets used in our experiments. For all datasets, we use a pre-trained language model \texttt{all-MiniLM-L6-v2} as a textual encoder to encode news headlines as the initial textual representations (dimension 384) of news nodes, which are then projected into a 64-dimensional space through a single-layer neural network.
\begin{itemize}[leftmargin=*]
    \item \textbf{Adressa}~\cite{gulla2017adressa}: A Norwegian news dataset released by NTNU and Adresseavisen, containing three months of user interaction logs (Jan--Mar 2017) with approximately 3M users, 50K articles, and 27M clicks. It provides complete timestamps for all user interactions, enabling fine-grained temporal modeling and stage-wise analysis.
    \item \textbf{MIND}~\cite{wu2020mind}: A large-scale English news dataset collected from MSN, spanning six weeks (Oct--Nov 2019) with approximately 1M users, 160K articles, and over 25M clicks.
    While the week-5 data contains fully timestamped labeled impressions, interactions of the weeks 1-4 data are only chronologically ordered without exact timestamps.
    
\end{itemize}

\subsubsection{Experimental Protocol}
Following prior work~\cite{wu2019neural,an2019neural,ko2025crown} in news recommendation, we also adopt a negative sampling strategy with a positive-to-negative ratio of 1:4 during training. This ratio is not treated as a tunable hyperparameter; it is fixed across all models to ensure consistency and fair comparison under the same cross-entropy ranking objective.

Note that our evaluation protocol is substantially more challenging than those used in many previous studies. Earlier studies typically evaluate on a much shorter temporal horizon (e.g., only the final day of week-5), whereas we use the entire week-5 as the test set. This broader temporal horizon introduces larger topical variation and stronger temporal drift, making candidate discrimination more difficult. As a result, traditional baselines without explicit mechanisms for modeling stage-wise transitions or long-range temporal evolution exhibit lower absolute AUC, while preserving stable relative performance ordering.

\subsubsection{Hardware and Software Environment}
We implemented our model using PyTorch 2.5 on Ubuntu 22.04.3, and all experiments were executed on a single NVIDIA L20 GPU with CUDA 12.x support. The proposed approach does not rely on multi-GPU parallelism or large-scale computing clusters, and can be efficiently trained on a single GPU.

\subsubsection{Hyperparameter Tuning and Sensitivity Analysis}
Hyperparameters are tuned using a grid-search strategy applied uniformly across all models to ensure fairness and reproducibility. Specifically, the learning rate is searched over $\{10^{-3}, 10^{-4}, 10^{-5}, 10^{-6}, 10^{-7}\}$, the weight decay over $\{10^{1}, 10^{0}, 10^{-1}, 10^{-2}, 10^{-3}\}$, and the temperature parameter over $\{10^{2}, 10^{1}, 10^{0}, 10^{-1}, 10^{-2}\}$.

The same search procedure is applied uniformly to all baselines to ensure fairness and consistency. Furthermore, all baseline implementations and parameter choices strictly follow the official code released by their original authors, ensuring reproducibility. Sensitivity analyses of the window size and loss-weight coefficients are reported in Figures~2 and~3, respectively. While real-world user histories can indeed be noisy, sparse, or bursty, the window size in our framework serves only as a coarse structural prior rather than a sensitive modeling hyperparameter. The actual irregular or abrupt shifts in interest are captured by our LPM module, which learns fine-grained dynamics within each stage. 

\subsubsection{Baselines}
To evaluate the effectiveness of our proposed approach, we compare it with a diverse set of competitive baselines, covering classical neural news recommenders, graph-based models, and recent temporal-aware methods:
  
\begin{itemize} [leftmargin=*]
    \item \textbf{NRMS (EMNLP 2019)}~\cite{wu2019neural}: A multi-head self-attention model that captures contextual dependencies among words and news for recommendation.  
    \item \textbf{NAML (IJCAI 2019)}~\cite{wu2019naml}:A multi-view attentive model that integrates title, body, and category information into hierarchical news representations.  
    \item \textbf{NPA (KDD 2019)}~\cite{wu2019npa}:A personalized attention model that dynamically assigns attention weights based on user preferences.
    \item \textbf{LightGCN (SIGIR 2020)}~\cite{he2020lightgcn}: A lightweight GCN model that propagates collaborative signals on the user–item graph without transformation or nonlinearity.
    \item \textbf{CNE-SUE (EMNLP 2021)}~\cite{mao2021neural}: A framework combining collaborative news encoding with a GCN-based user encoder through mutual representation learning.
    \item \textbf{TCCM (CIKM 2023)}~\cite{chen2023tccm}: A temporal context-aware model that explicitly captures the evolving dynamics of user preferences by constructing chronological subgraphs and leveraging contrastive learning. It serves as the most relevant baseline to our work, as it also aims to model time-dependent user interests.  
    \item \textbf{CROWN (WWW 2025)}~\cite{ko2025crown}: The most recent model, which leverages causal reasoning and disentangled representations to enhance user preference modeling. 
\end{itemize}


\subsubsection{Ablation Study Interpretation}
The substantial performance drop observed in ablation experiments highlights the complementary value of each module in our framework. Our evaluation setting introduces a longer temporal horizon and stronger distribution shifts than those considered in prior work, which amplifies the difficulty of the prediction task. Under such conditions, removing any component that models stage-wise transitions, long-term stability, or fine-grained temporal dynamics significantly degrades performance.

Although real-world news consumption is often noisy, sparse, and bursty, our method does not assume globally smooth preference evolution. Instead, the window size serves only as a coarse structural prior for stage segmentation, while irregular and abrupt interest shifts are explicitly captured by the proposed module that models fine-grained dynamics within each stage. This design explains why each component provides essential and non-redundant functionality, and why ablating any of them leads to a pronounced performance decline.

\end{document}